\documentclass{article}
\usepackage[utf8]{inputenc}
\usepackage[top=3cm,bottom=3cm,left=3cm,right=3cm]{geometry}

\usepackage[round]{natbib}
\usepackage{authblk}

\usepackage{bibentry}
\usepackage{pgfgantt}
\usepackage[perpage]{footmisc}
\usepackage{tabularx}
\usepackage{hyperref}
\usepackage{amssymb}
\usepackage{xcolor} 
\usepackage{hyperref}
\usepackage{cleveref}
\usepackage{comment}
\usepackage{booktabs}

\hypersetup{
    colorlinks=true,    
    urlcolor=blue,      
    linkcolor=blue,     
    citecolor=blue,     
    pdfborder={0 0 0}   
}

\begin{document}

\title{Shareholder Democracy with AI Representatives}

\author[1]{Suyash Fulay\thanks{These authors contributed equally.}}
\author[1,2]{Sercan Demir$^*$\thanks{Work done while at MIT.}}
\author[1]{Galen Hines-Pierce}
\author[3]{Hélène Landemore}
\author[1]{Michiel A. Bakker}

\affil[1]{Massachusetts Institute of Technology, Cambridge, MA, USA}
\affil[2]{University of Cologne, Cologne, Germany}
\affil[3]{Yale University, New Haven, CT, USA}

\date{}

\maketitle

\bigskip
\begin{abstract}
A large share of retail investors hold public equities through mutual funds, yet lack adequate control over these investments. Indeed, mutual funds concentrate voting power in the hands of a few asset managers. These managers vote on behalf of shareholders despite having limited insight into their individual preferences, leaving them exposed to growing political and regulatory pressures, particularly amid rising shareholder activism. Pass-through voting has been proposed as a way to empower retail investors and provide asset managers with clearer guidance, but it faces challenges such as low participation rates and the difficulty of capturing highly individualized shareholder preferences for each specific vote. Randomly selected assemblies of shareholders, or ``investor assemblies,'' have also been proposed as more representative proxies than asset managers. As a third alternative, we propose artificial intelligence (AI) enabled representatives trained on individual shareholder preferences to act as proxies and vote on their behalf. Over time, these models could not only predict how retail investors would vote at any given moment but also how they might vote if they had significantly more time, knowledge, and resources to evaluate each proposal, leading to better overall decision-making. We argue that shareholder democracy offers a compelling real-world test bed for AI-enabled representation, providing valuable insights into both the potential benefits and risks of this approach more generally.
\end{abstract}

\section*{Introduction}
Ensuring effective shareholder representation has been a topic of scholarly interest for decades \citep{BhagatBrickley1984, BethelGillan2002}.
In many ways, it provides a textbook example of collective decision-making: a large number of shareholders with heterogeneous preferences must choose among a set of limited options \citep{Yermack2010}.
Yet, adequately representing shareholder preferences is far from straightforward.
Investors in mutual funds and similar vehicles typically rely on asset managers to vote on their behalf, often with minimal direct input from the shareholders themselves \citep{Bainbridge2006}. Recently, large asset managers have faced increasing criticism for not addressing the heterogeneous preferences of shareholders, and there is a growing demand to give shareholders a stronger voice.\footnote{See \href{https://professional.ft.com/en-gb/blog/investors-demand-greater-esg-voting-rights/}{Investors demand greater ESG voting rights}, \textit{Financial Times.}}
However, since shareholder interests are heterogeneous and complex, including both financial and non-financial objectives (e.g., ESG considerations), asset managers face a complex problem representing them and cannot realistically account for the diverse and often conflicting preferences of shareholders in the complex landscape of corporate voting. Additionally, simply collecting votes from retail shareholders and voting accordingly has a variety of issues. Shareholders who hold individual stakes in companies face a classic collective action problem: any one voter’s share is usually too small to justify the time and effort needed to analyze extensive proxy materials, leaving many disengaged \citep{Zingales2024}, ultimately leaving critical governance decisions in the hands of the unrepresentative few.\footnote{See \citet{Zingales2024} who provide an excellent discussion of the shareholder representation problem.}
Against this backdrop, the central question emerges: \textit{How can individual shareholders be adequately represented at scale?}

One low-tech solution put forward by critics of the current system is to put in place a form of democratic representation inspired by citizens' assemblies, called ``investor assemblies'' \citep{Zingales2024}. Citizens' assemblies are bodies of a few hundred randomly selected citizens (at most) brought together to deliberate about important political issues and entrusted with making policy recommendations. They are now a well-established way for electoral systems to address their legitimacy deficit and increase their representativeness and responsiveness. Similarly, investor assemblies are proposed by their advocates as deliberative bodies of randomly selected investors entrusted to make recommendations on \textit{value-values} trade-offs that asset managers are currently struggling with.
While this solution is intriguing, it is arguably costly in various ways, including in terms of time investment for the participants. It is also unknown how accurate the representation achieved by such assemblies would be given the self-selection bias in who accepts to participate in such meetings. Finally, the risk of capture of such investor assemblies by powerful economic actors is also a concern.
Instead, or perhaps in addition to such human-dependent solutions, we turn to artificial intelligence (AI) as an alternative. Building on recent developments in AI, we propose large language model (LLM) representatives as a cheap, instantaneous, scalable, and incorruptible avenue to address the long-standing challenges in effective shareholder representation. LLM representatives are models that learn individual human behavior and act as representatives in digital environments \citep{jarrett2023language}. Inspired by recent work on implementing LLM representatives in collective decision-making, we propose shareholder representation as an ideal testbed to test LLM representatives in a practical, real-world setting \citep{jarrett2023language}. 
Specifically, we propose leveraging LLM representatives to represent individual human preferences in the context of shareholder democracy for the following reasons. 
First, LLM representatives could effectively serve as proxies for individual shareholders, capturing and conveying their preferences with high accuracy. Recent studies demonstrate that LLMs can infer and model human preferences across various domains \citep{goli_singh_2024, li2023elicitinghumanpreferenceslanguage, piriyakulkij2024activepreferenceinferenceusing} and accurately forecast voting behavior on novel policy statements based on prior decisions \citep{small2023opportunitiesrisksllmsscalable, Argyle_Busby_Fulda_Gubler_Rytting_Wingate_2023}. 
Second, as LLMs continue to improve, they may assist or even make decisions on behalf of retail shareholders \textit{as if} the investor had significantly more time and knowledge to evaluate shareholder proposals and their implications \citep{Yudkowsky2004CEV}.

The use of LLM representatives has the potential to democratize shareholder decision-making while maintaining scalability and efficiency, giving investors a meaningful voice without overwhelming the system. However, this approach is not without risks. AI representation may not predict preferences with equal accuracy across all groups \citep{whose_opinions}, potentially exacerbating disparities in representation. Moreover, it could lead to further disengagement among retail shareholders if they simply delegate all decision-making power to their AI representatives. More generally, some researchers have pointed out the risks of moving towards an ``algorithmic democracy'' wherein most decisions are offloaded to AI agents that purportedly represent citizens \citep{Garcia-Marza2024}. Over-reliance on AI representation could lead to further disengagement with the decision-making process and, in the long term, a reduced capability to understand and decide on important governance issues.  To mitigate these concerns, a robust oversight mechanism and communication of the representative's actions would be essential. Finally, this solution remains mostly aggregative as opposed to a deliberative one. LLM representatives are meant to track or simulate the views of atomized shareholders if they bothered to vote and did so in a way that is coherent with their claimed preferences. They act as delegates for the shareholders, in contrast to the investor assemblies, which act more as ``trustees'' with a degree of independence from the shareholders they represent. While there is some work that shows promise in using LLM representation in more deliberative settings \citep{jarrett2023language,ashkinaze2024plurals}, it remains an open question whether LLMs could also simulate the outcome of a structured deliberation between shareholders, which would create a cohesive group with an articulated, common view for their collective position. Until that question is resolved, representation via LLM is perhaps best seen as a prelude or a complement to more deliberative approaches, such as investor assemblies.

In this manuscript, we aim to make several contributions. First, drawing from existing approaches to shareholder representation, we propose a set of desiderata for an effective system. We then explore how AI-driven shareholder representation could function, along with the potential risks and limitations of this approach. Finally, we outline a research agenda to empirically evaluate the ability of LLMs to provide robust shareholder representation. More broadly, we argue that shareholder representation presents a compelling real-world use case for assessing both the potential and the limitations of LLM-driven decision-making.

\section*{Current Solutions}

In response to growing calls for increased shareholder representation, major asset managers like BlackRock, Vanguard, and State Street have introduced pass-through voting programs that delegate proxy voting power directly to shareholders \citep{BlackRockVotingChoice, Vanguard2023, StateStreet2023}. These initiatives, launched as early as 2021, have been implemented through structured voting frameworks provided by proxy advisors, firms that develop voting guidelines and make recommendations on how shares should be voted in corporate elections. Rather than casting votes on individual proposals, retail investors participate by selecting from predefined voting policies such as standard corporate governance approaches, ESG-focused strategies, or faith-based guidelines designed and administered by these proxy advisory firms \citep{proxy, Hu2025CustomProxy}.

This approach, branded as \textit{Voting Choice} or similar by the asset managers, gives shareholders a more direct role in corporate governance. However, several practical challenges remain. First, participation has been low. In Vanguard’s pilot program, for example, only 2\% of eligible retail investors opted in \citep{vanguard2025investorchoice}. Second, while the menu of policies has expanded over time, the structure still requires investors to choose from a limited set of bundled options. Even if more granular customization were available, most retail investors lack the time, resources, and expertise to engage deeply with each policy.

This model resembles a form of representative democracy: investors select the voting policy that best aligns with their values, much like voters choose elected officials. However, this approach shares some of the same limitations, namely, the need to accept imperfect bundles and the difficulty of staying informed and responsive over time \citep{FischSchwartz2023, Kobussen2024}.

Another approach proposed by \cite{Zingales2024} are the previously mentioned ``investor assemblies.'' These assemblies would recruit a small subset of shareholders (about 150) to discuss the various ethical and moral trade-offs associated with corporate initiatives and draft a set of guidelines for fund managers to follow. Since citizen assemblies have been used in the past to resolve diverse and contentious political issues such as electoral reform, marriage equality, abortion, end-of-life issues, climate justice, youth homelessness, or urban planning \citep{delib}, this is a promising approach to giving shareholders a voice. Additionally, deliberation can lead to higher-quality decisions, as participants have the time to learn from experts and hear a variety of perspectives. However, although representatives will be drawn from a pool of shareholders, only a small subset (at most hundreds of representatives out of millions of shareholders) will actually participate in the deliberation (which is true of all citizens' assemblies). Thus, participation is inherently limited, and the nuanced preferences of a significant portion of shareholders may not be fully represented. 

%%\section*{What are the trade-offs of different approaches to shareholder democracy?}
\section*{Desiderata of Shareholder Democracy}
Each approach to shareholder representation comes with its own set of advantages and drawbacks. For instance, pass-through voting grants retail investors more influence over decision-making, but it lacks scalability and may not always result in well-informed choices. Here, we outline key desiderata for an effective system of shareholder representation, drawing from theories of representative, deliberative, and direct democracy. These principles, inspired by the strengths and weaknesses of existing approaches, provide a framework for evaluating current methods and identifying their shortcomings. Moreover, they serve as a foundation for designing an AI-driven system of shareholder representation.

\begin{enumerate}
    \item \textbf{Individual Preference Representation} – Captures the extent to which an individual shareholder’s specific values and priorities are reflected in the final voting outcome. A system that forces shareholders to ``bundle'' their preferences (such as a steward model) may be inferior to one that captures nuanced preferences (such as pass-through voting). This is analogous to some of the trade-offs between direct \citep{altman2010direct} and representative \citep{urbinati2006representative} democracy.
    
    \item \textbf{Voting Preference Fidelity} – Assesses how accurately the voting mechanism captures and executes the preferences of shareholders. Even if a system aims to represent individual preferences, it may introduce distortions due to biases in preference elicitation, algorithmic limitations, or influence from external factors. 
    
    \item \textbf{Preference Quality} - Whether the system elicits preferences as if shareholders have learned about and reasoned through the proposal for a reasonable period of time. Pass-through voting, for example, may not allow participants to become as informed about the issue and the various perspectives as an investor assembly. This is where some proponents of deliberative democracy (often called "epistemic democrats") argue that their approach can lead to better decision-making \citep{delib}, with methods such as citizens' assemblies or deliberative polling \citep{fishkin2012deliberative} used to elicit more informed poll responses.
    
    \item \textbf{Scalability and Cost Efficiency} – Evaluates the feasibility of implementing the system at scale, considering both financial and time costs for mutual funds and retail investors. A system that demands active participation from millions of investors may be infeasible, whereas a more automated or representative approach could reduce costs but risk oversimplifying preferences.
    
    \item \textbf{Shareholder Engagement and Education} – Captures how well the system encourages retail investors to understand and participate in corporate governance decisions. A system that requires minimal input may be efficient but could disengage investors from the process, while a more deliberative approach may enhance awareness but require significant time and effort.
\end{enumerate}
Many of these considerations are not unique to shareholder representation. Some are fundamental questions that have a rich history in the political science literature, such as the trade-offs between direct and representative democracy on the one hand, and aggregative and deliberative approaches on the other. However, in this work, we attempted to ground our desiderata in the context of shareholder representation.

\section*{Are LLM-Mediated Shareholder Representatives a Feasible Approach?}

Consider a retail investor who generally supports climate-focused corporate initiatives but makes exceptions for industries in which they work or have specific financial interests. In a pass-through voting system, the investor would have the opportunity to express these nuanced preferences. However, in practice, most retail investors lack the time or willingness to engage with each policy and the hundreds of shareholder proposals presented each year. As a result, pass-through voting—despite offering high fidelity and individual preference representation—suffers from low participation rates and remains impractical at scale. Additionally, even if this retail investor did vote, their decision under time constraints may differ from their decision if they had more time and resources to study the issue. It may also differ from the decision they would reach in the context of a deliberation with other shareholders about their common interests and values.

An alternative approach, investor assemblies, could establish general voting guidelines for fund managers, reducing the burden on individual shareholders. This method improves cost efficiency and ensures well-reasoned, deliberative decision-making at the group level, as deliberative processes tend to produce thoughtful consensus respectful of and informed by minority objections and occasional dissent \citep{fournier2011citizens}. However, it may fail to capture the specific priorities of individual investors fully.

To bridge this gap, we propose AI voting representatives (henceforth \emph{AI representatives}) as a complementary solution. These are AI proxies trained on an investor’s stated preferences that could automate voting decisions, maintaining individual representation while ensuring scalability. Moreover, as LLMs improve in modeling personal preferences, they could go beyond simply mirroring an investor’s immediate choices. Instead, they could predict how an investor would vote if they had the time and resources to study each issue thoroughly. This concept aligns with \textit{extrapolated volition}, where AI systems are designed not just to reflect our present preferences but rather the choices we would make if we were more informed, rational, or experienced \citep{Yudkowsky2004CEV}.
In theory, this approach could approximate some of the benefits of investor assemblies, such as more informed, reasoned decision-making, while retaining the personalization and scalability of digital representation. However, whether AI representatives can capture the emergent properties of group deliberation remains an open question.

 \subsection*{Current Approaches to AI Representation}
 LLMs have shown that they are capable of predicting the opinions of various groups \citep{Argyle_Busby_Fulda_Gubler_Rytting_Wingate_2023}. Some work shows that the models can be better at predicting the response of majority versus minority groups \citep{whose_opinions} and can be prone to caricaturing these groups \citep{cheng-etal-2023-compost}. However, when given more personalized context, LLMs are more accurate at modeling individual survey responses and do not exhibit the same level of bias \citep{park2024generativeagentsimulations1000}.

 Given that LLMs have shown relatively high fidelity in predicting individual preferences, several recent works have attempted to investigate how well they predict policy preferences. \citet{small2023opportunitiesrisksllmsscalable} found that given previous votes on policy statements, LLMs were accurate and well-calibrated at predicting votes on unseen statements. \citet{jarrett2023language} also propose a framework for understanding how language models can act as representatives in collective decision-making and experimentally show that language models can be fine-tuned to reflect individual preferences. Given these results, we propose a way to use LLM-powered representatives to proxy shareholder preferences at scale.
\subsection*{Proposed Implementation for Shareholder Democracy}

\subsubsection*{Creating and Aggregating Shareholder Preferences}

A mutual fund consists of individual shareholders who contribute capital. Each shareholder $i$ invests an amount $s_i$, and their voting power is proportional to their total contribution:

$$
v_i = \frac{s_i}{\sum_{j=0}^{n}s_j}
$$
When the fund must decide on a proposal $p_k$ for company $k$, it can either approve, disapprove, or allocate votes in some proportion. To model shareholder preferences, we define a policy function:

$$
\pi_i: (x_i, p_k) \Rightarrow P(y_{ik} = 1 | x_i, p_k)
$$

where:
\begin{itemize}
    \item $x_i$ represents shareholder $i$’s preferences.
    \item $P(y_{ik} = 1 | x_i, p_k)$ is the probability that they support proposal $p_k$, where $y_{ik} = 1$ denotes approval and $y_{ik} = 0$ denotes disapproval.
\end{itemize}
$P(y_{ik} = 1 | x_i, p_k)$ can either be rounded to discretize each individual's vote or it can be left as a probability prior to the aggregation step.
\subsubsection*{Using LLMs to Estimate Shareholder Preferences}

LLMs can be leveraged to estimate these individual-level probabilities in different ways:

\begin{enumerate}
    \item \textbf{Prompting-Based Approach:}  
    Following \citet{whose_opinions}, \citet{Argyle_Busby_Fulda_Gubler_Rytting_Wingate_2023}, \citet{small2023opportunitiesrisksllmsscalable}, and \citet{park2024generativeagentsimulations1000}, we assume a shared policy function $\pi_i$ across all shareholders. Differences in individual voting behavior arise solely from variations in $x_i$. The LLM is prompted to determine whether shareholder $i$ would approve $p_k$ based on $x_i$. To obtain a probabilistic estimate, one could either extract token-level probabilities from the model’s output or directly prompt the LLM to provide a confidence score.

    \item \textbf{Fine-Tuning-Based Approach:}  
    Alternatively, \citet{jarrett2023language} demonstrated that fine-tuning LLMs on individual preference data enhances prediction accuracy. While fine-tuning could increase accuracy, it might make the model less general (e.g., if the fine-tuning data contains only decisions related to ESG, it might fail to generalize to other domains).
\end{enumerate}
We can validate the accuracy of these methods by collecting ground truth votes on existing shareholder proposals and seeing whether these methods can predict votes with sufficient fidelity. Using these individual-level predictions, the fund can estimate overall shareholder support for proposal $p_k$ by computing the weighted sum:

$$
\sum_{i=0}^{n} v_i \pi_i(x_i, p_k)
$$
This aggregate measure reflects the fund's investors' estimated stance on the proposal, incorporating the preferences of all shareholders in proportion to their investment. Funds could then decide whether to vote their shares proportionally or simply with the majority.

\subsubsection*{Collecting Shareholder Input}

To effectively train the policy or guide LLM prompts, shareholder preferences should be gathered both at the outset of their interaction with the system and on a rolling basis. This includes periodic updates, such as when shareholders revise their views or want to express input ahead of specific meetings. We recommend collecting a combination of general background information and specific policy preferences. If the fund anticipates key themes in upcoming proposals (e.g., climate policy), it should proactively solicit shareholder input on those topics. In addition to multiple-choice questions, open-ended prompts can help capture the reasoning behind shareholder views. This richer input allows the policy to better extrapolate and align with their expressed values.

\citet{park2024generativeagentsimulations1000} found that long-form interviews allowed for more accurate, unbiased predictions of survey responses, and we expect a similar effect when modeling shareholder preferences.
Ultimately, there will be a trade-off between the time shareholders invest in ``teaching'' their AI representative and the fidelity with which that representative captures their perspectives. Striking the right balance will be crucial to ensuring both efficiency and accuracy in shareholder-driven decision-making.
\subsubsection*{Shareholder Control of AI Representatives}
A fundamental aspect of effective representation is the ability to authorize and hold one’s representative accountable \citep{Dovi2018}. To ensure transparency and alignment with shareholder interests, AI representatives should provide real-time updates on their voting decisions. Ideally, shareholders would receive notifications detailing not only the decision made but also the reasoning behind it—leveraging LLMs’ capacity to generate human-readable explanations.
Beyond passive oversight, shareholders should also have the ability to actively refine their AI representatives over time. By offering additional context, correcting misinterpretations, or adjusting preferences based on evolving views, shareholders can iteratively improve their representatives’ alignment with their interests. This creates a dynamic feedback loop: shareholders supply data to train their AI representatives, the representatives cast votes on their behalf and justify their choices, and shareholders, in turn, refine their representatives' decision-making process. A similar iterative refinement has been used to create policy consensus statements with LLMs \citep{habermas}.

\section*{Risks, Limitations and Mitigations}
AI-driven shareholder voting can improve efficiency, scalability, and data-driven decision-making. Yet it raises significant concerns around accountability, systemic risk, and regulatory oversight. Despite these risks, AI has the potential to address persistent governance issues, particularly chronic low participation, where small but motivated groups disproportionately influence outcomes.

\subsection*{Who’s Accountable When AI Votes?}

AI proxy voting operates within uncertain legal frameworks regarding fiduciary responsibility. \citet{kroll2017accountable} note that algorithmic decisions lack contestability, complicating accountability when shareholder intent is misrepresented. Additionally, current financial regulations, historically designed for human decision-making and traditional financial intermediaries, face challenges adapting to automated, algorithm-driven financial innovations—a regulatory gap likely relevant to future AI-driven shareholder voting systems \citep{brummer2019fintech}. As \cite{garrett2025ai} argues, deploying AI in highly regulated contexts necessitates robust interpretability and procedural safeguards to meet due process requirements. He advocates for transparent, ``glass box" AI systems that clearly disclose decision-making criteria, allowing stakeholders meaningful opportunities to contest and correct errors. This emphasis parallels regulatory concerns in shareholder voting, where algorithmic opacity could complicate fiduciary accountability and oversight. Given that shareholder governance operates within stringent legal frameworks, Garrett's emphasis on interpretability and reliability highlights essential considerations for ensuring AI-driven voting complies with existing procedural standards and regulatory expectations \citep{garrett2025ai}.

Chronic non-participation in governance further complicates accountability. Most retail shareholders abstain from voting, enabling activist investors and proxy advisors to wield disproportionate influence \citep{cii2024proxyvoting}. Similarly, in democratic elections, low voter turnout magnifies the political impact of small, highly motivated groups \citep{olson1965logic}. AI technologies could potentially facilitate broader shareholder participation by simplifying voting processes and offering personalized decision-making support in democratic contexts. However, these systems must balance increased accessibility with stringent requirements for transparency and oversight to mitigate risks of algorithmic manipulation or capture.

\subsection*{Balancing AI Benefits with Risks of Systemic Bias}

AI governance systems inherently reflect the biases embedded in training data and institutional practices. For example, \citet{matsusaka2024robo} highlight that robot-voting can reinforce existing governance biases rather than promote deliberative pluralism. Similarly, \citet{kirilenko2013moore} illustrate the risks of systemic instability arising from algorithmic decision-making, as exemplified by the 2010 ``Flash Crash," where automated trading intensified market volatility through herding behavior. These examples underscore the importance of careful system design in AI-driven governance. While AI has the potential to activate passive shareholders and introduce structured, neutral information in democratic contexts, achieving these benefits requires deliberate safeguards against unintended algorithmic convergence and reinforcement of existing majority preferences.

\subsection*{Gradual Disempowerment and the Retail Investor Paradox}

The increasing use of AI to represent human preferences or make decisions on our behalf carries a significant risk of ``gradual disempowerment,''  \citep{kulveit2025gradual}. When AI representatives take on complex tasks and decision-making, human oversight can diminish, and our understanding of underlying processes may fade. Over time, this can erode our genuine influence over outcomes, even if the AI initially acts in our interest. While this is a critical concern for AI representation broadly, shareholder voting presents a nuanced scenario. Given that retail investor participation is currently often minimal, AI representatives trained on their preferences could paradoxically \textit{enhance} their influence by providing a systematic voice.

Furthermore, such systems might \textit{boost} retail investor engagement. AI representatives could explain their voting rationale, especially for unexpected decisions, and highlight proposals aligned with an investor’s core values, helping them focus on pertinent issues rather than being overwhelmed. This AI-facilitated interaction could cultivate a more informed and active shareholder base.

Crucially, this specific application of AI also offers a valuable testbed to empirically measure the complex dynamics of human trust in AI-driven decision-making. As AI takes on representative roles, understanding how investors' trust evolves, based on perceived performance, transparency of reasoning, and alignment with their values, is important. Observing how explanations build or erode trust, and whether trust leads to deeper engagement or passive delegation, can provide critical data. These empirical insights into trust dynamics are essential for anticipating and navigating the challenges of maintaining meaningful human agency as AI becomes more deeply embedded across various societal domains. Thus, while broader risks warrant caution, AI-enabled voting in this context might uniquely improve retail shareholder influence and engagement, while simultaneously serving as a practical laboratory for studying human-AI trust.

\subsection*{Infrastructure Rather than Reactive Regulation}

Rather than relying on reactive regulatory solutions, governance frameworks require robust, standardized infrastructure to ensure AI-driven decision-making remains transparent, accountable, and contestable. \citet{chan2025aiagents} propose a digital public infrastructure modeled after internet protocols, emphasizing transparency, resilience, and decentralized control. In corporate governance, such infrastructure would enable transparent and auditable AI-generated decisions, allow customizable AI models aligned with diverse governance philosophies, and ensure decentralization to prevent concentrated algorithmic control. Similarly, democratic systems would benefit from infrastructure that enhances voter participation through transparent, contestable election tools, decentralizes AI-driven decision-making, and provides resistance to algorithmic manipulation.

\subsection*{Looking Forward}

Integrating AI into governance requires prioritizing transparency, accountability, and broad-based participation. Achieving these goals necessitates proactive, deliberate system design that maximizes the potential benefits of AI while minimizing inherent risks. Ultimately, AI governance should aim to foster greater inclusivity and pluralism, avoiding the concentration of decision-making authority in the hands of a few.

\section*{Research Directions}

The framework proposed for AI-mediated shareholder representation opens several avenues for future inquiry, which is essential for robustly validating and refining such systems. A critical first step involves empirically understanding the nuances of shareholder preferences. This could be achieved through diverse data collection methods, ranging from analyzing publicly available shareholder proposals and voting records to conducting qualitative interviews and controlled studies with retail investors. Such research would aim to capture not only stated preferences but also the underlying values and reasoning, providing a richer dataset for training and evaluating AI representatives.

Further research should focus on the modeling techniques themselves. Comparative studies of different LLM architectures and prompting strategies, versus fine-tuning approaches, are needed to determine optimal methods for predicting shareholder votes with high fidelity and fairness across diverse investor profiles and proposal types. Investigating how to effectively model, evaluate, and represent extrapolated preferences (how an investor \textit{might} vote if more informed or after deliberation) without imposing a singular ``correct'' preference will be a key challenge. This involves exploring how AI can present potential future selves or reasoned outcomes back to the user for their consideration and validation.

The design of the human-AI interface is another crucial area. Research is needed on how best to present AI-generated voting explanations to shareholders in a way that fosters trust and understanding without undue persuasion. Exploring interactive feedback loops, where shareholders can easily query, correct, and refine their AI representative's understanding, will be vital for maintaining alignment and user agency. This includes studying how different levels of AI autonomy affect shareholder engagement and the perceived legitimacy of the system.

Beyond individual representation, the potential for AI to facilitate collective sense-making and agenda-setting warrants exploration. Could LLM ensembles, representing diverse shareholder factions, deliberate on complex issues and generate novel proposals or identify common ground? This moves beyond mere preference aggregation to using AI as a tool for constructive dialogue and policy co-creation within the shareholder ecosystem.

Finally, real-world pilot programs in collaboration with asset managers will be necessary. Such deployments would allow for the assessment of these systems' scalability, practical adoption challenges, and their actual impact on shareholder engagement, decision quality, and overall corporate governance dynamics. These pilots would also be crucial for observing long-term effects on user trust, potential over-reliance, and the ethical implications of AI representatives operating in high-stakes financial environments. Longitudinal studies within these pilots could track how shareholder understanding and participation evolve with sustained AI interaction.

Advancing AI-mediated shareholder representation requires a multi-faceted approach, combining technical development with rigorous empirical validation and ethical scrutiny to ensure these systems genuinely enhance, rather than diminish, democratic principles in corporate governance.

\section*{Discussion}

The concept of LLM representatives in shareholder democracy serves as a valuable testbed for broader LLM-driven representation in decision-making processes. Shareholder governance, characterized by dispersed and often disengaged retail investors, presents a microcosm of larger societal decision-making challenges. By examining the effectiveness of LLM representatives in this context, we gain insights into the broader potential and risks of LLM representation in democratic institutions, policy-making, and collective decision-making at scale.
One of the most promising aspects of LLM-driven representation is its ability to capture and execute individual preferences at scale. Unlike traditional democratic processes, where citizens delegate decisions to elected officials or predefined policy bundles, LLM representatives can provide highly personalized decision-making that reflects an individual's nuanced priorities. This capability could extend beyond shareholder voting into areas such as political representation, legislative decision-making, and participatory governance. In this sense, shareholder democracy provides an ideal controlled environment to test how LLMs might function as an intermediary in larger democratic systems.
While LLM-representatives offer several advantages, they pose challenges, including bias in preference modeling and risks to democratic engagement. If not appropriately managed, biases may reinforce inequalities, and reliance on LLMs could erode human participation.
Shareholder democracy serves as a practical testing ground for LLM representation in decision-making, offering valuable insights into scalability, engagement, and ethical considerations. The lessons learned from this domain can inform broader applications of LLM representatives in governance, law, and policy-making. However, responsible deployment requires ongoing scrutiny to balance efficiency with democratic integrity, ensuring that LLM representatives enhance rather than undermine participatory decision-making structures.

\section*{Acknowledgements}
We thank Nadya Malenko for helpful comments.

\bibliography{bibliography}

\end{document}